\documentclass{aa}
\usepackage{natbib}
\usepackage{graphicx}
\usepackage{amssymb}
\usepackage{color}
\usepackage{amsmath}
\usepackage{threeparttable}
\usepackage{tabularx}
\usepackage{caption}

\newcommand{\bl}[1]{\mbox{\boldmath$ #1 $}}

\begin{document}

\title{Formation of freely floating sub-stellar objects via close encounters}

\author{Eduard I. Vorobyov \inst{1,2}, Maria E. Steinrueck \inst{3}, Vardan Elbakyan \inst{2}, and Manuel Guedel \inst{4}}
\authorrunning{Vorobyov et al.}

\institute{Institute of Fluid Mechanics and Heat Transfer, TU Wien, 1060, Vienna, Austria; 
\and
Research Institute of Physics, Southern Federal University, Stachki Ave. 194, Rostov-on-Don, 
344090 Russia; 
\and
Lunar and Planetary Laboratory, University of Arizona, Tucson, AZ 85721, USA
\and 
University of Vienna, Department of Astrophysics,  Vienna, 1180, Austria
}

\abstract 
{}
{We numerically studied close encounters between a young stellar system hosting a massive, gravitationally
fragmenting disk and  an intruder diskless star with the purpose to determine the evolution 
of fragments that have formed in the disk prior to the encounter.}   
{Numerical hydrodynamics simulations in the non-inertial frame of reference of the host star were employed
to simulate the prograde and retrograde  co-planar encounters. The initial configuration of the 
target system (star plus disk) was obtained via a separate numerical simulation featuring 
the gravitational collapse of a solar-mass pre-stellar core.}
{We found that close encounters can lead to the ejection of fragments that have formed in 
the disk of the target prior to collision. In particular, prograde encounters are more efficient 
in ejecting the fragments than the retrograde encounters. The masses of ejected fragments
are in the brown-dwarf mass regime. They also carry away
an appreciable amount of gas in their gravitational radius of influence, implying that these
objects may possess extended disks or envelopes, as also suggested by Thies et al. (2015).
Close encounters can also lead to the ejection of entire spiral arms, followed by
fragmentation and formation of freely-floating objects straddling the planetary mass limit. However,
numerical simulations with a higher resolution are needed to confirm this finding. 
  }
  {}
 
\authorrunning{Vorobyov et al.}
\titlerunning{Formation of freely floating objects}
\maketitle

\section{Introduction}

Most stars do not form in isolation but in associations or clusters, and
encounters between young stars in the clusters are a natural outcome of the star formation
process caused by its turbulent nature \citep[e.g.][]{Kroupa1995,Lada2003,Bate2009}. 
Encounters can significantly influence the evolution of protostellar disks, 
which has been confirmed by numerical studies combining 
cluster simulations with simulations of individual encounters
\citep[e.g.][]{Pfalzner2008a,Pfalzner2008b,Rosotti2014}. Among the
expected effects are the truncation of disks and enhanced accretion rates.
In addition, various more detailed simulations of individual encounters have
been performed \citep[e.g.][]{Shen2006, Shen2010,Forgan2009,Forgan2010, Thies2010,Munoz2015}, 
investigating the
role of encounters in FU Orionis outbursts, brown dwarf formation, disk
instability, truncation of disks, and binary capture. In particular, 
\citet{Thies2010} demonstrated that a close encounter can trigger gravitational
fragmentation in the disk of a target which is otherwise stable to fragmentation.

None of these studies, however, considered encounters which involved a
disk that was already prone to fragmentation before the encounter. Such disks
are expected to exist in the early embedded stages of evolution 
\citep[e.g.][]{Machida2010,Vor2011,Tsukamoto2013,Dunham2014}. 
Moreover, the likelihood of encounters 
with young systems with still massive and extended disk is expected to be higher than 
with their old counterparts 
because the cluster members spread around with age due to dynamical scattering.  
For instance, \citet{Thies2010} estimated that an average $0.5~M_\odot$ star in an 
Orion-type cluster can experience
an encounter with a periastron radius below several hundred~AU with a probability of $\sim 10\%$, 
while being young and  hosting a massive and extended disk.

Encounters involving fragmenting disks can be more complex, as the approaching
star does not only interact gravitationally with the disk, but also with the fragments. This
might lead to perturbations of the fragment trajectories, leading to scattering or even ejection 
of fragments formed in the disk before the encounter. This may provide a source of freely floating
brown-dwarf and even planetary-mass objects, in addition to those that may be ejected from the disk
via internal multi-body interactions \citep{Stamatellos2009, Bate2009, BV2012,Vor2016}.

In this paper, the collision scenario is explored with numerical hydrodynamics
simulations featuring a close, co-planar passage of a Solar-type star with a massive, fragmenting disk around a $0.63~M_\odot$ star. To generate realistic disk configurations, we ran our model from the
gravitational collapse of a pre-stellar core until the central star and disk were formed  and initiated the collision with the intruder when the age of the system was 0.42~Myr and the mass of the disk was of $0.25~M_\odot$. At this time instance, the disk already featured several fragments formed via gravitational
fragmentation. 

The layout of the paper is as follows. The model description and initial conditions are 
presented in Sect.~\ref{modeldesc}. The main results are presented in Sect.~\ref{results}.
The model caveats are discussed in Sect.~\ref{discuss}. The main findings are summarized in Sect.~\ref{conclude}.

\section{Model description}
\label{modeldesc}
In this section, the main features of the numerical model used for studying close
encounters with young protostellar disks are summarized.
To compute the formation and
early evolution of protostellar disks, we employed the numerical hydrodynamics 
code described in detail in \citet{VB2015}. This code was further modified to include 
the close passage of an intruder star represented by a gravitating point source.
We start our numerical simulations from the gravitational collapse of a prestellar cloud core,
continue into the embedded phase of star formation, during which a star and disk 
are formed, and terminate our simulations when less than 10\%
of the initial core mass remains in the envelope. 
To accelerate the computations, we use the thin-disk approximation \citep[for justification see][]{VB2010}.

To avoid too small time steps, we introduce a "sink cell" at
$r_{\rm sc} = 8.0$~AU. The sink cell is dynamically inactive; it contributes only to the total 
gravitational potential and secures a smooth behavior of the gravity force down to the
stellar surface. 
During the early stages of the core collapse, we monitor the gas surface density in the sink cell and when its value exceeds a critical value for the transition from isothermal
to adiabatic evolution, we introduce a central point-mass star. In the subsequent evolution, 
90\% of the gas that flows to the sink cell is assumed to land on the growing star.
The other 10\% of the accreted gas is assumed to be carried
away with protostellar jets. When the in-spiralling material of the collapsing core hits the centrifugal barrier near the sink cell the protostellar disk starts to grow, occupying the inner part of the numerical polar grid.  In the embedded phase of disk evolution, the disk 
is exposed to intense mass-loading from the infalling envelope,
which increases the disk mass and triggers the onset of gravitational instability.
The disk becomes prone to fragment usually in the Class~I phase when the disk mass and size are the largest. In the early T Tauri phase
when the parental core nearly dissipates via accretion,
we launch the intruder towards the newly formed star plus disk system and further follow the 
dynamics of the encounter for several tens of thousand years.

\subsection{Disk hydrodynamics equations}
\label{two-D}
The equations of mass, momentum, and energy transport describing the dynamics of 
circumstellar disks in the thin-disk limit can be formulated as follows:
\begin{equation}
\label{cont}
\frac{{\partial \Sigma }}{{\partial t}} =  - \nabla  \cdot 
\left( \Sigma \bl{v} \right),  
\end{equation}
\begin{eqnarray}
\label{mom}
\frac{\partial}{\partial t} \left( \Sigma \bl{v} \right) + \nabla \cdot \left( \Sigma \bl{v}
\otimes \bl{v} \right) &=&  \Sigma \, \left( \bl{g}_\ast + \bl{g}_{\rm disk} -\bl{g}_{\rm ind}+ \bl{g}_{\rm intr} \right)  - \nonumber \\ &-&  \nabla {\cal P} + \nabla \cdot \mathbf{\Pi}, 
\end{eqnarray}
\begin{equation}
\label{energ}
\frac{\partial e}{\partial t} +\nabla \cdot \left( e \bl{v} \right) = -{\cal P} 
(\nabla \cdot \bl{v}) -\Lambda +\Gamma + 
\left(\nabla \bl{v}\right):\mathbf{\Pi}, 
\end{equation}
where $\Sigma$ is the mass surface density, $e$ is the internal energy per 
surface area, ${\cal P}$ is the vertically integrated gas pressure calculated 
via the ideal equation of state as ${\cal P}=(\gamma-1) e$ with $\gamma=7/5$, 
$\bl{v}=v_r \hat{\bl r}+ v_\phi \hat{\bl \phi}$ is the velocity in the
disk plane, and $\nabla=\hat{\bl r} \partial / \partial r + \hat{\bl \phi} r^{-1} 
\partial / \partial \phi $ is the gradient along the planar coordinates of the disk.
The gravitational accelerations in the disk plane, $\bl{g}_{\rm disk}$ and $\bl{g}_\ast$,
take into account self-gravity of the disk and the gravity of the central protostar when formed. 
Disk self-gravity is found  by solving for the Poisson integral: 
\begin{eqnarray} 
  \Phi(r,\phi) & = & - G \int_{r_{\rm sc}}^{r_{\rm out}} r^\prime dr^\prime 
                     \nonumber \\ 
      & &       \times \int_0^{2\pi} 
               \frac{\Sigma(r^\prime,\phi^\prime) d\phi^\prime} 
                    {\sqrt{{r^\prime}^2 + r^2 - 2 r r^\prime 
                       \cos(\phi^\prime - \phi) }}  \, ,
\end{eqnarray} 
where  $r_{\rm sc}$ and $r_{\rm out}$ are the radial positions of the computational inner and
outer boundaries. This integral is calculated using a FFT technique which applies 
the 2D Fourier convolution theorem for polar coordinates 
and allows for the non-periodic boundary conditions in the $r$-direction  by effectively
doubling the computation domain in this coordinate direction and filling it with zero densities 
\citep[see][Sect.\ 2.8]{BT1987}.

The evolution of isolated circumstellar disks is usually computed in the inertial frame of reference,
in which the position of the star is fixed in the coordinate center. This approach is justified for
as long as the gravitational influence of the disk on the star is negligible, 
which is true for low-mass, nearly-axisymmetric disks. However, the intruder will exert a strong 
gravitational pull on the target star so that it cannot be anymore considered as motionless.
In addition, young massive disks are often non-axisymmetric so that the superposition of 
gravity forces acting on the star from the disk may also be non-zero, causing the star to move (or wobble)
in response.  The grid-based methods applied to studying the disk dynamics often use 
cylindrical  or 
spherical coordinate systems, which are naturally centered on the star. 
However, the use of curvilinear coordinate systems with a singularity point in the coordinate
center makes it extremely difficult to calculate directly the stellar motion and its 
gravitational response to the disk and/or intruder.
To circumvent this problem,
a non-inertial frame of reference is often introduced, which moves with
the star (positioned in the coordinate center) in response to the net gravitational force 
of the disk and/or intruder. The resulting acceleration
of the star can be described as
\begin{equation}
\bl{g}_{\rm ind} = G \int \frac{dm (\bl{r'})}{r'^3} \, \bl{r'} + 
G \frac{M_{\rm intr}}{r_{\rm intr}^3} 
\bl{r}_{\rm intr},
\end{equation}
where $dm (\bl{r'})$ is a mass element in the disk with position vector $\bl{r'}$, $M_{\rm intr}$ is the mass of the intruder and $\bl{r}_{\rm intr}$ is the position vector of the intruder. The term that has to be added to the total gravitational acceleration thus is $-\bl{g}_{\rm ind}$. This term can be expressed as the derivative of a potential, the so-called indirect potential 
$\Phi_{\rm ind}=\bl{r} \cdot \bl{g}_{\rm ind}$;
the details of calculating $\Phi_{\rm ind}$ are provided in \citet{Regaly2017}. 
The details on how the gravitational acceleration from the intruder $\bl{g}_{\rm intr}$ 
is calculated will be provided in Section~\ref{sec:intruder}.

Turbulent viscosity is taken into account via the viscous stress tensor 
$\mathbf{\Pi}$, the expression for which can be found in \citet{VB2010}.
The kinematic viscosity needed to calculate the viscous stress tensor is found 
adopting the Shakura and Sunyaev parameterization \citep{SS1973}, so that $\nu=\alpha c_{\rm s} h$,
where $c_{\rm s}=\sqrt{\gamma {\cal P}/\Sigma}$ is the sound speed and $h$ is the disk scale height
calculated from the assumption of local hydrostatic balance in the disk.
In this work, we use a spatially and temporally constant $\alpha=10^{-2}$.


The radiative cooling per surface area $\Lambda$ in Equation~(\ref{mom}) is determined using
the diffusion approximation of the vertical radiation transport in
a one-zone model of the vertical disk structure \citep{Dong2016}
\begin{equation}
\Lambda=\frac{4\tau_{\rm P} \sigma T_{\rm mp}^4 }{1+2\tau_{\rm P} + 
{3 \over 2}\tau_{\rm R}\tau_{\rm P}},
\end{equation}
where $\tau_{\rm R}=\kappa_{\rm R} \Sigma_{1/2}$ 
and $\tau_{\rm P}=\kappa_{\rm P} \Sigma_{1/2}$ are the 
Rosseland and Planck optical depths to the disk midplane,  $\kappa_P$ and 
$\kappa_R$ the Planck and Rosseland mean opacities, $\Sigma_{1/2}=\Sigma/2$ 
the gas surface density from the disk surface to the midplane, and
$T_{\rm mp}={\cal P} \mu/\Sigma {\cal R}$ a form of the midplane gas temperature.

The heating function per surface area of the disk is expressed as
\begin{equation}
\Gamma=\frac{4\tau_{\rm P} \sigma T_{\rm irr}^4 }{1+2\tau_{\rm P} + 
{3 \over 2}\tau_{\rm R}\tau_{\rm P}},
\end{equation}
where $T_{\rm irr}$ is the irradiation temperature at the disk surface 
determined by the stellar and background black-body irradiation as
\begin{equation}
T_{\rm irr}^4=T_{\rm bg}^4+\frac{F_{\rm irr}(r)}{\sigma},
\label{fluxCS}
\end{equation}
where $T_{\rm bg}$ is the uniform background temperature (in our model set to the 
initial temperature of the natal cloud core)
and $F_{\rm irr}(r)$ is the radiation flux (energy per unit time per unit surface area) 
absorbed by the disk surface at radial distance 
$r$ from the central object. The latter quantity is calculated as 
\begin{equation}
F_{\rm irr}(r)= \frac{L_{\ast}}{4\pi r^2} \cos{\gamma_{\rm irr}},
\label{fluxF}
\end{equation}
where $\gamma_{\rm irr}$ is the incidence angle of 
radiation arriving at the disk surface at radial distance $r$. The incidence angle is calculated
using the disk surface curvature inferred from the radial profile of the  
disk vertical scale height at each time step \citep{VB2010}.
The stellar luminosity $L_\ast$ is the sum of the accretion luminosity
$L_{\ast,\rm accr}=G M_\ast\dot{M}/2R_\ast$ arising from the gravitational
energy of accreted gas and the photospheric luminosity $L_{\ast,\rm ph}$
due to gravitational compression and deuterium burning in the
stellar interior. The stellar mass $M_\ast$ and accretion rate onto the
star $\dot{M}$ are determined using the amount of gas passing through
the sink cell, while the stellar radius $R_\ast$ and $L_{\ast, \rm ph}$ are 
returned by the Lyon stellar evolution code \citep{Baraffe2010}. 

Equations (\ref{cont})--(\ref{energ}) are solved in the polar coordinates on a
numerical grid with $512\times512$ grid zones. The solution
procedure is similar in methodology to the ZEUS code \citep{SN1992} and 
is described in detail in \citet{VB2010}. We impose a free outflow boundary condition so that
the matter is allowed to flow out of the computational domain 
but is prevented from flowing in. Once the intruder is launched, 
the outer boundary condition is switched to the free inflow/outflow (and not only outflow) one. 
This is done to prevent artificial rarefaction of matter near the 
outer boundary in the non-inertial frame of reference moving with the target star.

\subsection{Description of the intruder}
\label{sec:intruder}
Because of the adopted thin-disk approximation, the motion of the intruder is restricted to the disk plane. Coplanar encounters are expected to have the highest impact on the disk \citep{Forgan2009} and our numerical code is best suited for this type of encounters. The equations of motion for the intruder moving in a combined gravitational potential of the target star and the disk, written as first order differential equations  in polar coordinates, are given by
\begin{eqnarray}
\dot r & = & v_r, \label{eqn:eom1}\\
\dot \varphi & = & \frac{v_\varphi}{r}, \\
\dot v_r & = & \frac{v_\varphi^2}{r} + a_r, \\ 
\dot v_\varphi & = & - \frac{v_\varphi v_r}{r} + a_\varphi, 
\label{eqn:eom4}
\end{eqnarray}
with the acceleration calculated as
\begin{equation}
\bl{a}= -\nabla(\Phi_\ast + \Phi_{\rm ind}) + \frac{1}{M_{\rm intr}} \sum_{j,k} \bl{F}_{j,k},
\end{equation} 
where $\bl{F}_{j,k}$ is the force the intruder experiences from a single grid cell, given in equation (\ref{eqn:intruderforce}), and the summation is performed over all grid cells.
To calculate the trajectory of the intruder, equations (\ref{eqn:eom1})-(\ref{eqn:eom4}) are solved using a Dormand-Prince method (a fifth-order Runge-Kutta method) with adaptive stepsize control \citep[see][chap. 17.2]{Press2007}. Several tests  were performed to demonstrate that the code is producing correct results \citep{Steinrueck}.

\subsubsection{Intruder-Disk interaction}
To calculate the influence of the intruder on the gas dynamics in the disk, the gravitational potential of the intruder is added to the total potential on each grid point in the hydrodynamics code. 
The gravitational potential of the intruder at a grid cell with indices $j$ and $k$ is given by 
\begin{equation}
\label{eqn:intruderpotential}
\Phi_{\rm intr} (R_{j,k}) =
\begin{cases}
        - G M_{\rm intr}/R_{j,k} & \text{for } R_{j,k}\ge r_{\rm s} \\
        \Phi_{\rm smooth} & \text{for } R_{j,k}< r_{\rm s},
\end{cases}
\end{equation}
where $r_{\rm s}$ is the smoothing radius and $\Phi_{\rm smooth}$ is given by equation (\ref{eqn:smoothedpotential}). The symbol $\bl{R}_{j,k} =  \bl{r}_{j,k} - \bl{r}_{intr} $ is used for the vector pointing from the intruder towards the center of the grid cell and the distance between the grid cell and the intruder is denoted by $R_{j,k}=\vert \bl{R}_{j,k} \vert$.

\subsubsection{Disk-Intruder interaction}
The force that the disk exerts onto the intruder is calculated by summing the contributions from each cell, treating the cells as point masses with mass $M_{j,k}$. The contribution from an individual cell is given by
\begin{equation}
\label{eqn:intruderforce}
\bl{F}_{j,k} =
\begin{cases}
 \displaystyle \frac{G M_{\rm intr} M_{j,k}}{R_{j,k}^3} \bl{R}_{j,k}   
 & \text{for } R_{j,k} \ge r_{\rm s} \\
\displaystyle \bl{F}_{j,k}^{(\rm smooth)} & \text{for } R_{j,k} < r_{\rm s},
\end{cases}
\end{equation}
with $\bl{F}_{j,k}^{(\rm smooth)}$ defined in equation (\ref{eqn:smoothedforce}).
As the summation is computationally expensive, the forces are updated only once per hydrodynamical timestep. 

\subsection{Gravitational softening}
\label{subsec:softening}
The gravitational forces between the gas and the intruder can become extremely strong if the intruder approaches one of the cell centers, where all mass is concentrated according to our convention. To avoid this unphysical behavior, the potential has to be modified for small distances. We 
use the same potential as in \citet{Klahr2006}, where the potential for distances smaller than a smoothing radius $r_{\rm s}$ is set to
\begin{equation}
\label{eqn:smoothedpotential}
\Phi_{\rm smooth}(R_{j,k}) = -GM_{\rm intr} \left( \frac{R_{j,k}^3}{r_{\rm s}^4} - 2 \frac{R_{j,k}^2}{r_{\rm
s}^3} + \frac{2}{r_{\rm s}} \right).
\end{equation}
At $r_s$ the smoothed potential and its first and second derivatives agree with the unaltered potential $-GM_{\rm intr}/R_{j,k}$ and its derivatives. The advantage of this choice is that it only modifies the potential for distances smaller than $r_{\rm s}$, unlike the common approach 
$\Phi(R_{j,k})=-GM/ \sqrt{R_{j,k}^2+r_{\rm s}^2}$.

The modified force that a single grid cell with a distance from the intruder smaller than 
$r_{\rm s}$ exerts on the intruder is then given by
\begin{equation}
\label{eqn:smoothedforce}
\bl{F}_{j,k}^{(\rm smooth)}= G M_{\rm intr} M_{j,k}
\left(\frac{4}{r_{\rm s}^3} - \frac{3 R_{j,k}}{r_{\rm s}^4} \right) \bl{R}_{j,k}.
\end{equation}

The choice of the smoothing radius $r_{\rm s}$ is not trivial. If it is too small, the gravitational interaction of the gas and the intruder will be overestimated, a too large $r_{\rm s}$ will underestimate the forces. As the smoothing is introduced to compensate discretization effects, it seems plausible that the smoothing radius has to be related to the cell size: 
\begin{equation}
r_{\rm s}=q_{\rm s} \cdot \min (\Delta r,r \Delta \phi), 
\end{equation} 
where $\Delta r$ and $\Delta \phi$ are the local cell spacings in the radial and azimuthal direction. Testing different smoothing radii shows that $q_{\rm s}=1$ results in unexpectedly strong interactions with the disk, altering the trajectory of the intruder more than expected, while there is relatively little difference between $q_{\rm s}=2$ and $q_{\rm s}=3$. We therefore choose $q_{\rm s}=3$ throughout the simulations.

\section{Initial conditions and the early disk evolution}
\label{sec:initialconditions}

To obtain a realistic initial disk configuration, a simulation of the collapse of a cloud core was run, similar to the simulations in \citet{VB2015}.  The initial surface density and angular velocity 
profiles of the core are as follows
\begin{eqnarray}
\Sigma(r) & = & {r_0 \Sigma_0 \over \sqrt{r^2+r_0^2}}\:, \\
\Omega(r) & = &2\Omega_0 \left( {r_0\over r}\right)^2 \left[\sqrt{1+\left({r\over r_0}\right)^2
} -1\right],
\label{ic}
\end{eqnarray}
where $\Sigma_0$ and $\Omega_0$ are the gas surface density and angular velocity 
at the center of the core. These profiles have a small near-uniform
central region of size $r_0$ and then transition to an $r^{-1}$ profile;
they are representative of a wide class of observations and theoretical models
\citep{Andre93,Dapp09}. For the chosen $r_0=2400$~AU, $\Sigma_0=5.16\times10^{-2}$~g~cm$^{-2}$, 
and $\Omega_0=1.2$~km~s$^{-1}$~pc$^{-1}$, the total mass of the core is $M_{\rm core}=1.08~M_\odot$ and the ratio of rotational to gravitational energy $\beta$ is $6.2 \times 10^{-3}$.

\begin{figure*}
\begin{centering}
\includegraphics[scale=0.9]{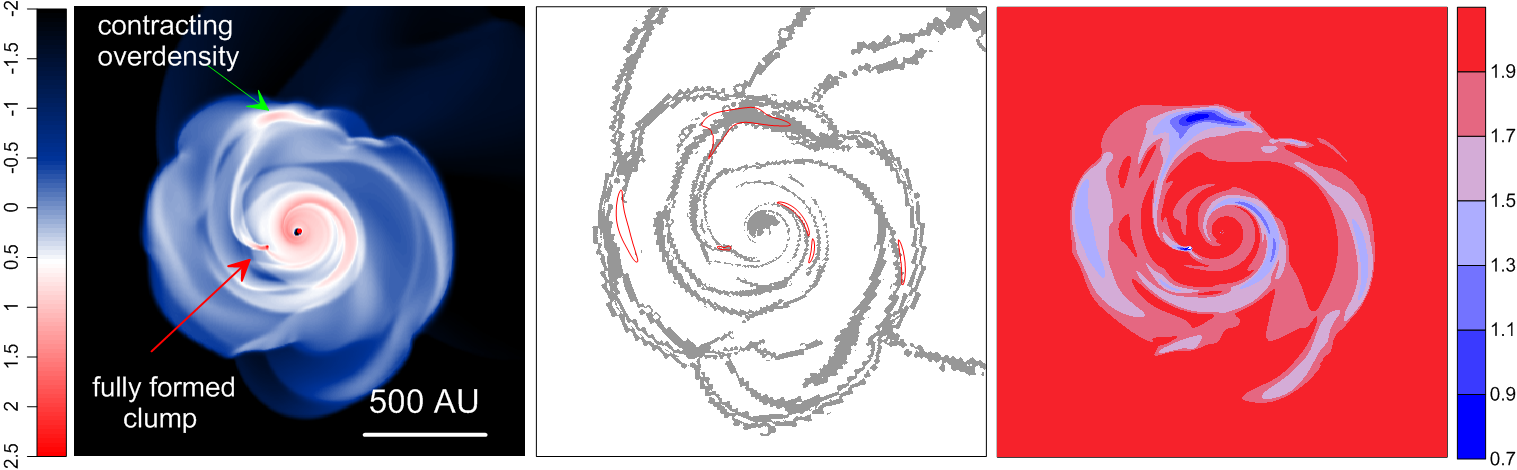} 
\par\end{centering}
\protect\protect\caption{{\bf Left}. Surface density of the circumstellar disk at the time at which the intruder is added to the simulation. Only the region inside 1000 AU from the central star is shown.
The green and red arrows show the forming and already formed fragments in the disk. {\bf Middle}. Red
contour lines show the regions with the Toomre $Q$-parameter smaller than unity, while the grey-shaded
area indicated the disk regions where the Gammie ${\cal G}$-parameter is smaller than 3. {\bf Right}.
Ratio $N_{\rm Jeans}$ (in log scale) of the Jeans length $R_{\rm J}$ to the maximum size of the local grid cell $\max(\Delta
r, r \Delta\phi)$. The minimum value of $N_{\rm Jeans}$ is found in the center of the 
contracting overdensity and is equal to 5.  In other parts of the contracting overdensity, 
$N_{\rm Jeans}$ is about 10, which  means the the Truelove criterion is fulfilled. }
\label{fig1}
\end{figure*}

We choose the disk configuration at $t=0.42$~Myr after the onset of the core collapse as the
initial setup for all simulations including the intruder. At this time,
the central star has a mass of $M_\ast=0.63~M_{\odot}$ and the disk mass is 0.25~$M_{\odot}$. 
The surface density inside 1000 AU from the star at this time is plotted in 
the left-hand-side panel of 
Figure \ref{fig1}. The disk extends to around 500~AU from the central star and 
is gravitationally unstable as is manifested by a well-developed spiral structure. 
There is one fragment present in the disk to the left of the central star (shown with the red arrow). At the top, at around 500 AU distance from the central star, an overdensity can be seen (shown with the green arrow). Soon afterwards it collapses and forms another fragment. 

To check if the the conditions in the disk fulfill the known fragmentation criteria, we plot in the
middle panel of Figure~\ref{fig1} the spatial distribution of the Toomre $Q$-parameter,
$Q=c_{\rm s}\Omega/\pi G \Sigma$ and the Gammie ${\cal G}$-parameter defined as ${\cal G}=t_{\rm cool}
\Omega$  \citep{Gammie2001}, where $t_{\rm cool}= e/(\Lambda-\Gamma)$ is the local cooling time. 
We note that if $\Lambda - \Gamma < 0$, cooling turns into heating and both the cooling time and the Gammie criterion become formally negative.  
The gravitationally 
unstable disk is expected to fragment if $Q<1.0$ and $0< {\cal G} < 3-5$. The numerical values
in both criteria may vary somewhat depending on physical conditions in the disk and resolution 
\citep{Meru2011, Takahashi2016,Boss2017}, but we have chosen the conservative values.   
The disk regions where the Q-parameter is smaller than unity are shown with the red contour lines, while
the regions where the ${\cal G}$-parameter is smaller than 3.0 are plotted with the grey color.
Clearly, the contracting overdensity fulfills both criteria. The same can be said about the already
existing fragment. However, there are several regions in the disk where the $Q$-parameter is smaller than unity, but the ${\cal G}$-parameter is greater than 3.0. These regions show no sign of fragmentation.

Finally, we check if the Truelove criterion \citep{Truelove1998}, stating that the Jeans length should be resolved by at least 4 grid cells in each coordinate direction, is fulfilled. In the thin-disk limit, the Jeans length is defined as $R_{\rm J}= \langle v^2 \rangle/\pi G \Sigma$, 
where $ \langle v^2 \rangle$ is the velocity dispersion
in the disk plane \citep{Vor2013}.
The right-hand-side panel in Figure~\ref{fig1} presents the ratio 
$N_{\rm Jeans}=R_{\rm J}/\max(\Delta r, r \Delta
\phi)$ (in log scale)  calculated for every grid cell in the disk, 
where $\Delta r$ and $r \Delta \phi$ are the grid spacings in the radial 
and azimuthal direction. The minimum of 
$N_{\rm Jeans}$ is found at the centers of the forming and already
formed fragments. In the most relevant case of the forming fragment (we do not
expect a secondary fragmentation in the already formed fragment), 
the minimum value of $N_{\rm Jeans}$ is 5 at the center of the forming fragment and increases to 10 around the fragment, meaning that the Jeans
length is resolved. The ratio $N_{\rm Jeans}$ is greater than 10 in the rest of the disk.

\section{Results}
\label{results}

\begin{table*}
\renewcommand{\arraystretch}{1.2}
\center
\caption{Overview of the simulations}
\label{table1}
\begin{tabular}{ccccccc }
\hline\hline

 Model & $v_{r,0}$ (km/s) & $v_{\phi,0}$ (km/s) & $\phi_0$ (deg) & expected periastron\tnote{a} (AU) & actual periastron (AU) & Fragment ejected? \\

   \hline
   P1           & -1.8 &  0.5 &  90 & 473 & 412 & yes \\
   P2           & -2.1 &  0.7 &  90 & 704 & 659 & yes \\
   P3           & -1.9 &  0.6 &  90 & 606 & 556 & yes \\
   P4           & -1.7 & 0.45 &  90 & 410 & 338 & yes (2) \\
   P5           & -1.8 &  0.5 & 180 & 473 & 391 & yes (2) \\
   P6           & -1.8 &  0.5 & 270 & 473 & 467 & yes (2) \\
   \hline
   R1           & -1.8 &  -0.5 &  90 & 473 & 466 & yes \\
   R2           & -2.1 &  -0.7 &  90 & 704 & 700 & no \\
   R3           & -1.9 &  -0.6 &  90 & 606 & 601 & no \\
   R4           & -1.7 & -0.45 &  90 & 410 & 397 & no \\
   R5           & -1.8 &  -0.5 & 270 & 473 & 423 & no \\
   R6           & -1.8 &  -0.5 &   0 & 473 & 384 & no \\
   \hline

\end{tabular}
\center{The columns show, from left to right, the model identifier (the names of prograde models start with P, the ones of retrograde models with R), radial and azimuthal components of the initial velocity of the intruder $v_{r,0}$ and $v_{\phi,0}$, the initial azimuthal position of the intruder $\phi_0$, expected  (calculated analytically treating the disk-star system as a point mass) and actual periastron radius and whether fragments have been ejected during the simulation.}
\end{table*}

In this section, we presents the results of numerical simulations showing the outcome of collisions
between a point-mass intruder star and a target star hosting a massive, gravitationally unstable disk
that has already formed several fragments.
We considered several models of encounters with a mass of the intruder star $M_{\rm intr}=1.2~M_\odot$.
The intruder is added to the simulation at $t=0.42$~Myr  
at a distance of 3000 AU, far enough to avoid any effects from the sudden appearance of the intruder on the disk of the target. Table~\ref{table1} presents the overview of our simulations. In total,
we have run 12 models with different initial velocities of the intruder: 6 out 12 models are characterized by the prograde trajectory of the intruder with respect to the disk of the target and the 
rest have retrograde trajectories. All encounters are coplanar with the disk of the target.
We have found several interesting phenomena related to collisions of the intruder star with the disk
of the target which we describe in detail below.

\begin{figure*}
\begin{centering}
\includegraphics[scale=0.6]{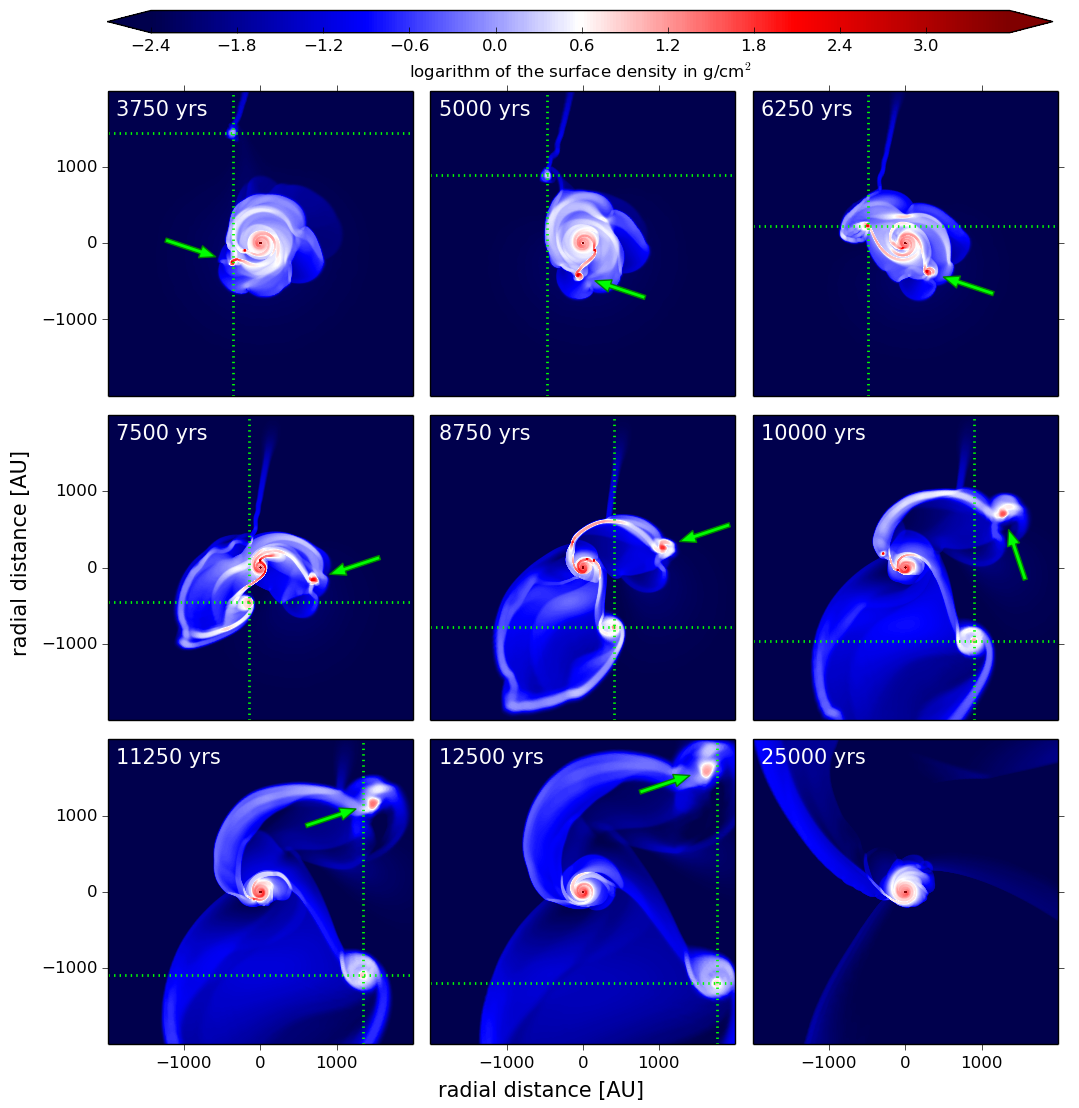} 
\par\end{centering}
\protect\protect\caption{A series of simulation snapshots showing the ejection of a fragment from the disk through three-body interaction with the intruder and the central star in Model~P1. The intersection of the dotted green lines marks the position of the intruder. The position of the fragment that is ejected is indicated with an arrow. The last panel shows the disk long after the encounter.}
\label{fig2}
\end{figure*}

\subsection{Ejection of existing fragments}

In the majority of the prograde models, one of the fragments is later 
ejected from the disk due to three-body interaction between the intruder, the target star and the fragment
itself. Figure \ref{fig2} illustrates this process with a series of simulation snapshots from Model P1. Each panel shows the surface density in g~cm$^{-2}$ on a logarithmic scale in a 4000 AU x 4000 AU box centered around the target star. In this figure, as well as in all other figures showing simulation snapshots, the position of the intruder is indicated by the intersection of the dotted lines. 
The position of the ejected fragment is highlighted by an arrow. The time shown is elapsed 
since the launch of the intruder.

 In the initial stages, the approaching star is clearly visible thanks to a gas tail which 
 the intruder leaves in its wake when passing through the remnant envelope surrounding 
 the target system. The closest approach occurs
 at around $t=6500$~yr with a periastron distance of 412~AU. By this time, the inner, less 
 massive fragment (shown
 by the red arrow in Figure~\ref{fig1}) has migrated into the inner disk regions and dispersed by the
 action of tidal torques, but the outer, more massive fragment formed from a contracting overdensity
 (shown by the green arrow in Figure~\ref{fig1})  has survived the encounter and 
 gained angular momentum via the three-body gravitational interaction between the intruder star, 
 target star+disk system and the fragment itself.

\begin{figure*}
\begin{centering}
\includegraphics[scale=0.8]{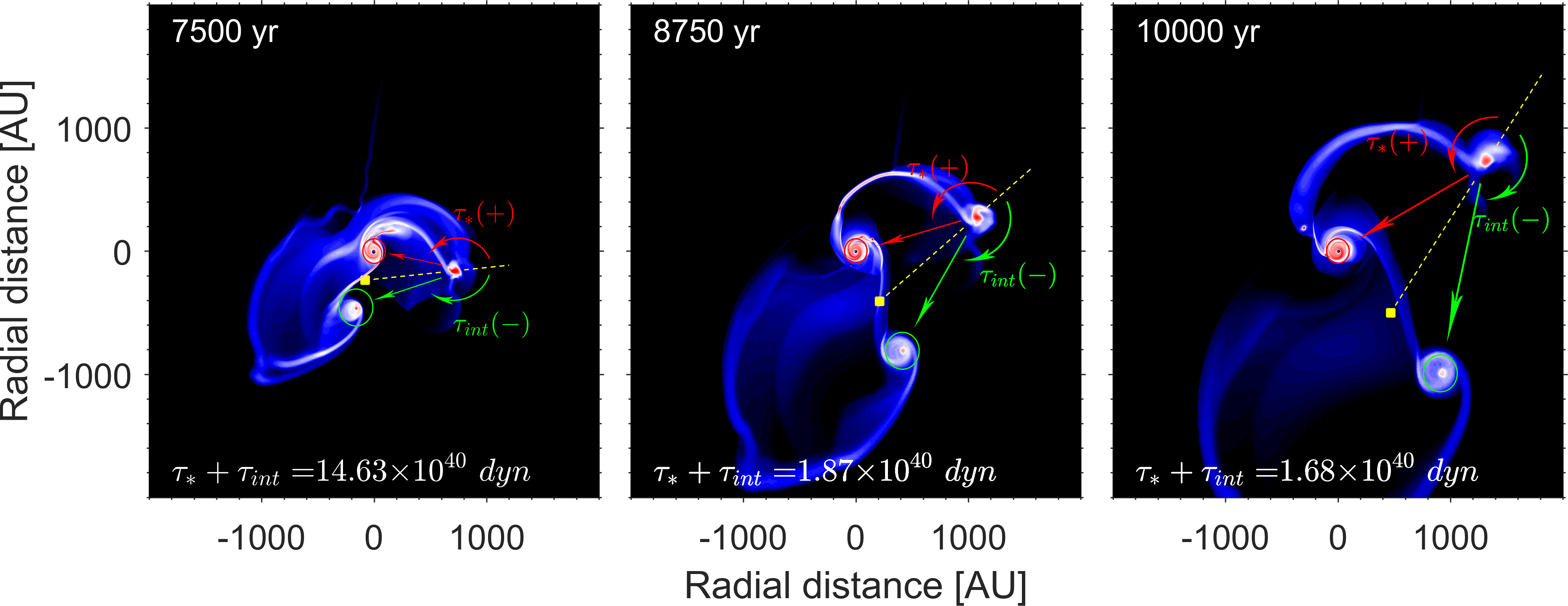} 
\par\end{centering}
\protect\protect\caption{Schematic illustration of the gravitational forces and torques acting on the ejected clump from the target star and the intruder. Three consecutive times similar to those in the
middle row of Figure~\ref{fig2} are chosen. The clump orbits the host star in the counter-clockwise
direction.  The red and green arrows show the gravity 
forces acting on the ejected clump from the target star and intruder, respectively. The yellow 
square marks the center of mass of the host star
and the intruder (including their disks outlined by the red and green circles) 
and the dashed yellow line represents the lever arm applied to calculate the torques. 
The gravitational torques acting on the 
ejected clump from the host star $\tau_\ast$ (positive) and the intruder $\tau_{\rm int}$ (negative) are shown by the curved
red and green arrows, respectively. The sum of the torques acting on the clump is always positive and its numerical value
is provided in the bottom of each panel. }
\label{fig2a}
\end{figure*}

Figure~\ref{fig2a} illustrates the ejection mechanism and shows schematically the gravitational forces
and torques acting on the ejected clump from the target star and the intruder 
at three consecutive times corresponding to the middle
row in Figure~\ref{fig2}. In particular, the red and green arrows show the gravity 
forces acting on
the ejected clump from the target star and intruder, respectively. When calculating these forces, 
we considered the total masses of the target star and the intruder including the gas disks around 
them (outlined by the red and green circles, respectively). The yellow square marks the center of mass of the host star
and the intruder (including their disks) and the dashed yellow line represents the arm. 
The gravitational torques acting on the 
ejected clump from the host star $\tau_\ast$ and the intruder $\tau_{\rm int}$ are shown by the curved
red and green arrows, respectively. Because the clump orbits the host star in the counter-clockwise direction, 
$\tau_\ast$ is positive and $\tau_{\rm int}$ is negative. The sum of the two torques is always positive,
as indicated in each panel, meaning that the clump gains angular momentum. Once its kinetic energy becomes
greater than the potential gravitational energy (see Table~2), it permanently leaves the
system.


To calculate the mass of the ejected fragments, we employed the fragment tracking algorithm described
in detail in \citet{Vor2013}. Namely, two conditions were used: 1) the fragment must be pressure supported,
with a negative pressure gradient with respect to the
center of the fragment; and 2) the fragment must be kept together
by gravity, with the potential well being deepest at the center of
the fragment. The resulting masses of the ejected fragments, along with their other 
characteristics, are presented in Table~\ref{tab2}.

\begin{table*}
\renewcommand{\arraystretch}{1.2}
\center
\caption{Properties of ejected fragments formed inside the disk}
\label{tab2}
\begin{tabular}{cccccc}
\hline\hline

   Model &  $M_{\rm frag}$ (M$_{\rm Jup}$) & $M_{\rm Hill}$ (M$_{\rm Jup}$) & $v_{\rm s}$ (km/s) 
   & $v_{\rm i}$ (km/s) & $ E_{\rm kin}/\vert E_{\rm grav} \vert$  \\

    \hline
   P1           & 24 & 45 & 1.67 & 2.20 & 3.32 (CM)  \\
   P2           & 21 & 40 & 1.18 & 2.80 & 2.93 (S)   \\
   P3           & 23 & 41 & 1.37 & 2.56 & 4.54 (S)   \\
   P4           & 25 & 44 & 1.46 & 1.65 & 1.75 (CM)  \\
   P6           & 10 & 23 & 1.88 & 2.15 & 2.92 (CM)  \\
   R1           & 10 & 25 & 2.13 & 3.33 & 7.32 (CM)  \\
   \hline

\end{tabular}
\center{The columns show, from left to right, the fragment mass $M_{\rm frag}$, the mass located inside the fragment's Hill radius $M_{\rm Hill}$, the velocity of the fragment relative to the disk central star $v_{\rm s}$ and the intruder $v_{\rm i}$, and the ratio of kinetic to potential energy in the reference frame in which this ratio is the lowest with the reference frame indicated in brackets (CM=center of mass system, S=central star). }
\end{table*}

With only slight variations between different models, the mass of the ejected fragment is 20--25 Jupiter masses. 
An exception is Model~P6 with 10~M$_{\rm Jup}$, as the intruder passes the fragment relatively close, such that the fragment loses part of its mass via tidal stripping.
There is an additional amount of gas with 40--45 M$_{\rm Jup}$ (23 M$_{\rm Jup}$ in Model~P6) located inside the Hill radius of the fragment. This mass may be accreted by the fragment and its final mass is likely to be in the brown dwarf regime.  Thus, the fragment resembles 
a proto-brown dwarf. A fraction of the material in the Hill radius may form an accretion disk, as observed around some young brown dwarfs \citep[e.g.][]{Riaz2016}. The ratio of kinetic to gravitational 
potential energy of the 
ejected fragment $E_{\rm kin}/ \vert E_{\rm grav} \vert$  in Table~\ref{tab2} is 
greater than unity, implying that  we witness true ejection and not gravitational scatter to some eccentric orbit around the target star. Table~\ref{tab2} also indicates that the ejection velocities are 
a few km~s$^{-1}$, 
which is consistent with the mean velocities of brown dwarfs in young star forming regions
\citep[e.g.][]{Kroupa2003,Joergens2006}. For instance, Kroupa \& Bouvier write in their abstract ``It is shown here that these results can be understood if brown dwarfs are produced as ejected 
embryos 
with a dispersion of ejection velocities of about 2.0~km~s$^{-1}$, and if the number of ejected 
embryos is about one per four stars born in Taurus-Auriga and the Orion Nebular cluster''.

\begin{figure}
\begin{centering}
\resizebox{\hsize}{!}{\includegraphics{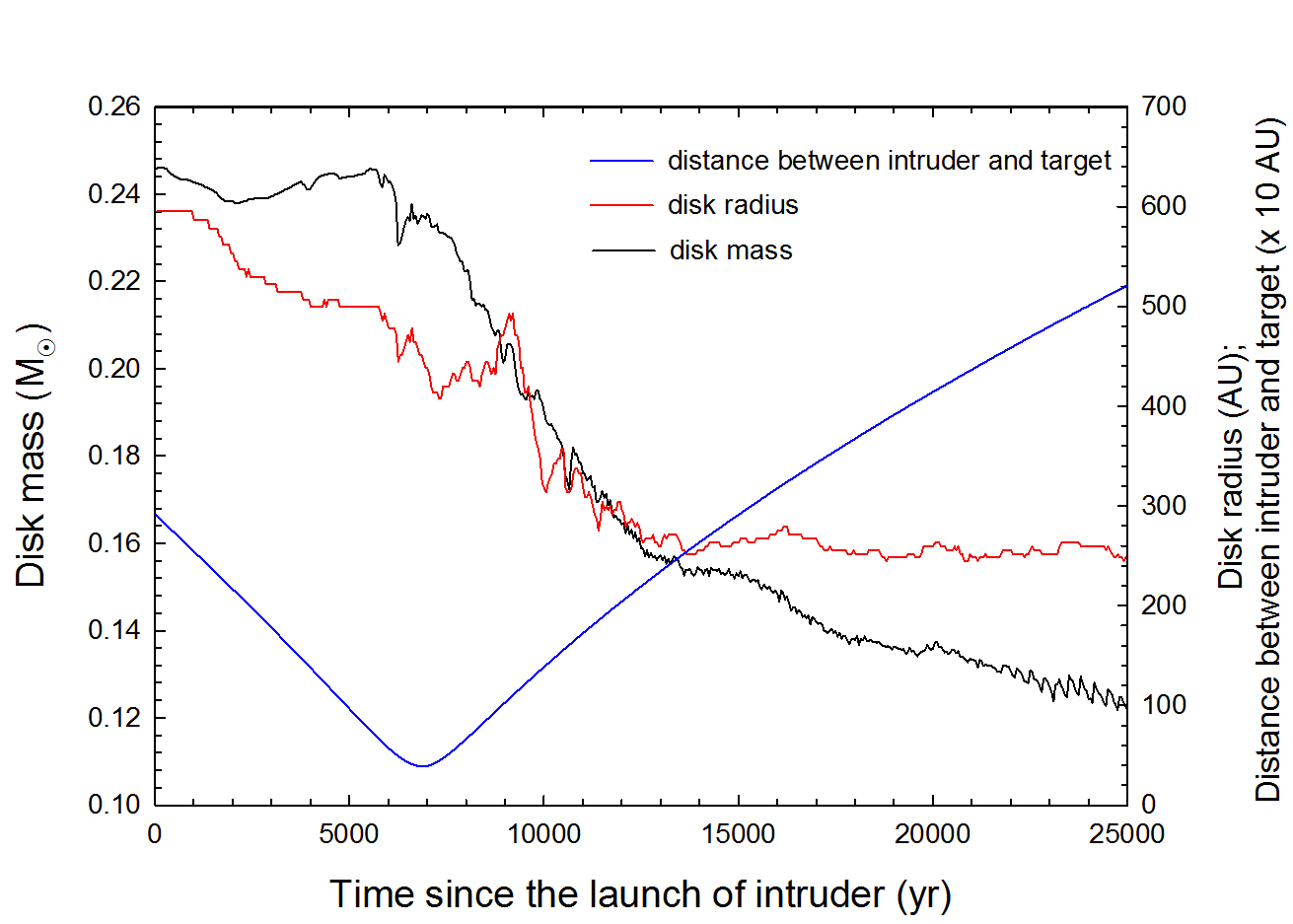}}
\par\end{centering}
\centering{}\protect\protect\protect\caption{Variations in the disk mass and radius (the black and red
lines) as a function of time elapsed since the launch of the intruder star. The blue line shows the
distance of the intruder from the target star.}
\label{fig2b}
\end{figure}

\begin{figure*}
\begin{centering}
\includegraphics[scale=0.6]{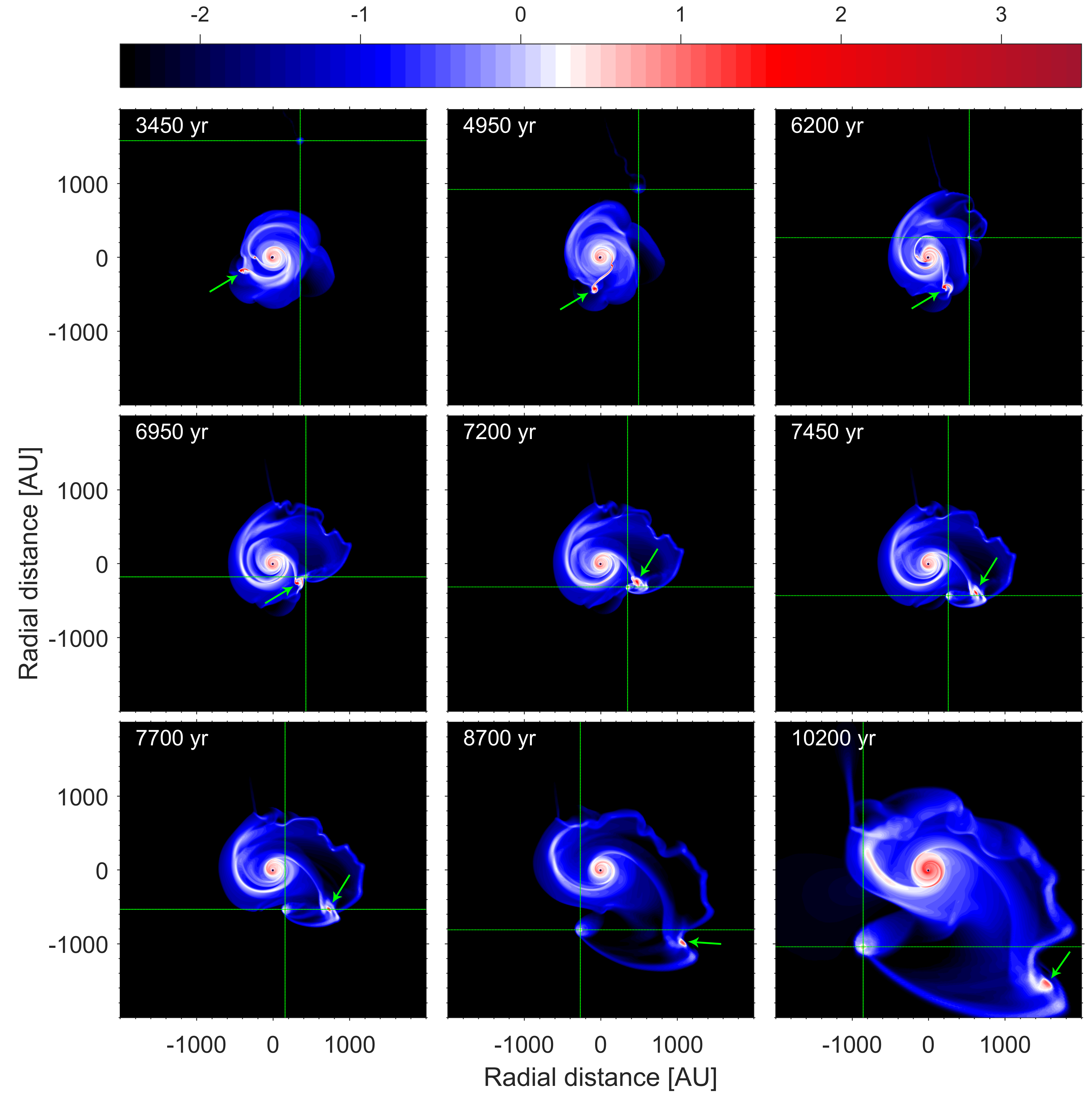} 
\par\end{centering}
\protect\protect\caption{A series of simulation snapshots showing the ejection of a fragment from the disk through three-body interaction with the intruder and the central star in Model~R1. The intersection of the dotted green lines marks the position of the intruder. The position of the fragment that is ejected is indicated with an arrow. The last panel shows the disk long after the encounter.}
\label{fig2bb}
\end{figure*}

An important effect that was also reported in previous simulations with close encounters 
\citep[e.g.][]{Pfalzner2008a} is truncation of the disk around the target star. This phenomenon 
is illustrated in 
Figure~\ref{fig2b} showing the time evolution of the disk mass and radius (the black and red lines, respectively) along with radial distance of the intruder from the target star (the blue line) in model~P1.
Clearly, the collision of the intruder with the disk around the target star causes a significant reduction
in the disk mass and size. While before the collision, the disk mass and radius were $0.24~M_\odot$
and 550~AU, 18000~yr after the closest approach the respective values are $0.16~M_\odot$ and $100$~AU.
The truncated disk still shows signs of gravitational instability as manifested by flocculent spiral
arms, but is stable against gravitational fragmentation.
It is also evident that the intruder  captures and retains some material from the target disk. This material later forms a compact rotating gaseous disk around the intruder. 
We also note that accretion of disk material onto the intruding star has been studied by 
\citet{Thies2011} and \citet{Kroupa2017} who showed that this process may be important for the 
formation of misaligned hot Jupiters.

In most of the models in which the intruder has a retrograde trajectory, no
ejections are observed. This may be related to the fact that in retrograde collisions 
the net angular momentum of the involved constituents: the intruder, the target, and 
the target disk (including fragments) is smaller as compared to the case with prograde collisions, making
ejections a less likely outcome. Figure~\ref{fig2bb} (model R1), however, presents a  spectacular 
exception. In this model,
the intruder passes the fragment (shown by the green arrow) very closely and catapults the fragment
out of the system by means of its gravitational pull. 
The ejection process is illustrated in Figure~\ref{fig2bb}, showing
the surface density at nine consecutive time instances. In the top row, the intruder
approaches the fragment. The closest approach to the fragment, with a distance of $\approx60$~AU,
happens in the middle row. As can be seen, the fragment
gets highly distorted due to tidal forces and loses part of its mass. The bottom row shows how the fragment
leaves the system.
The ejection in retrograde collisions requires therefore a close encounter, which is expected to be
a rare event.
For comparison, the distance between the intruder star and the ejected fragment in the 
prograde model~P1 at the closest approach to each other is about 1000~AU.

 \begin{table*}
\renewcommand{\arraystretch}{1.2}
\center
\caption{Properties of ejected fragments formed in ejected spiral arms}
\label{table3}
\begin{tabular}{cccccc }
\hline\hline

   Model &  $M_{\rm frag}$ (M$_{\rm Jup}$) & $M_{\rm Hill}$ (M$_{\rm Jup}$) & $v_{\rm s}$ (km/s) 
   & $v_{\rm i}$ (km/s) & $ E_{\rm kin}/\vert E_{\rm grav} \vert$  \\

    \hline
   P4         &  9 & 13 & 1.20 & 2.08 & 3.03 (S)   \\
   P5         &  9 & 21\tnote{$\ast$} & 1.39 & 2.47 & 2.94 (S)   \\
              &  5 & 14\tnote{$\ast$}& 1.22 & 2.26 & 2.06 (S)   \\
   P6         & 10 & 24 & 1.46 & 2.21 & 3.25 (CM)  \\
   \hline

\end{tabular}
\caption*{        The columns show, from left to right, the fragment mass $M_{\rm frag}$, the mass located inside the fragment's Hill radius $M_{\rm Hill}$, the velocity of the fragment relative to the disk central star $v_{\rm s}$ and the intruder $v_{\rm i}$, and  the ratio of kinetic to potential energy in the reference frame in which this ratio is the lowest with the reference frame indicated in brackets (CM=center of mass system, S=central star).}    
\end{table*}

\subsection{Ejection of spiral arms followed by their gravitational fragmentation}
\label{spiral}

\begin{figure*}
\begin{centering}
\includegraphics[scale=0.6]{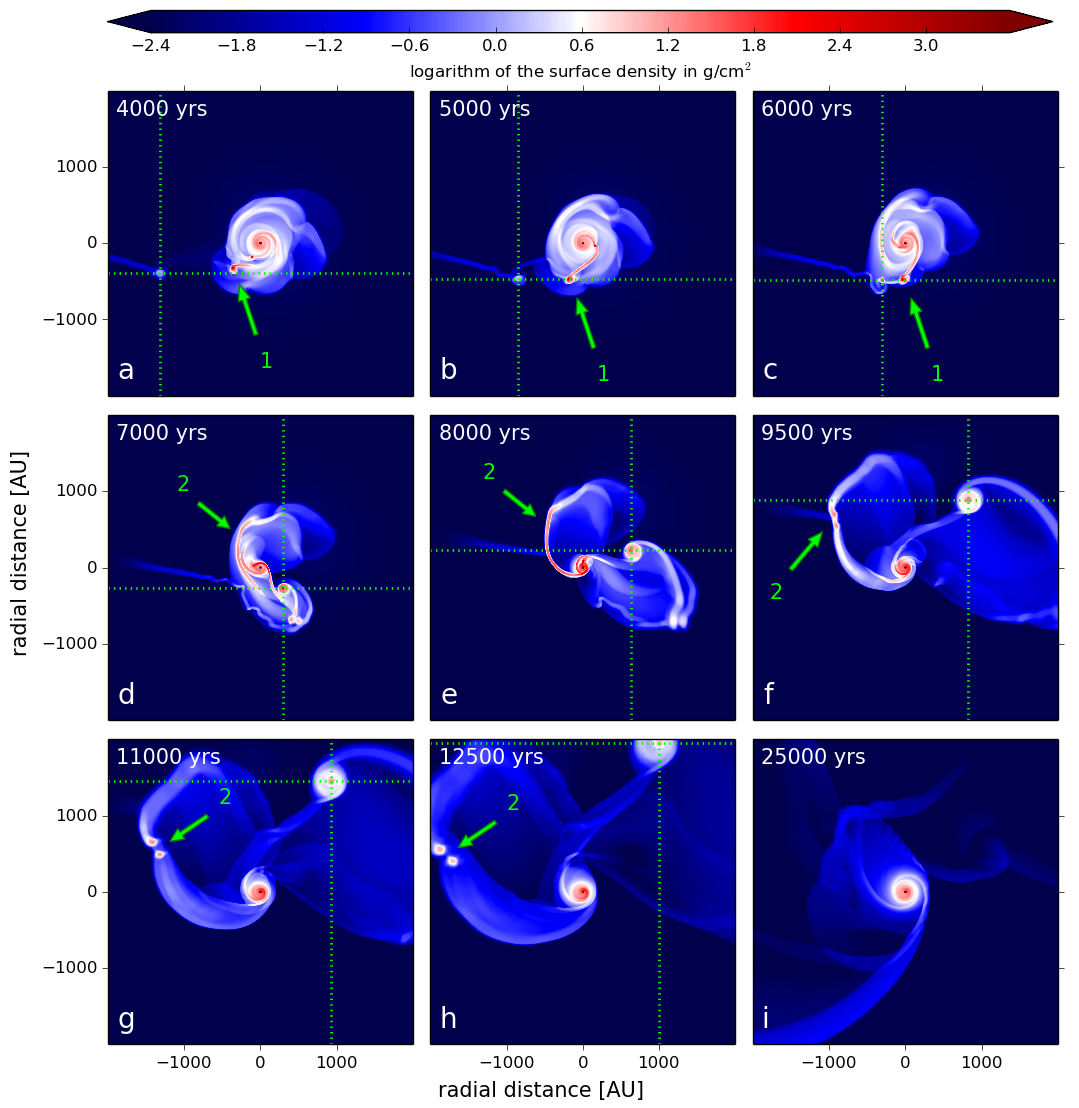} 
\par\end{centering}
\protect\protect\caption{A series of simulation snapshots showing the formation of two
fragments in the ejected spiral arm in Model~P5. Again, the position of the intruder is
indicated by the intersection of the dotted green lines. Arrow~1 points to the pre-existing fragment,
while Arrow~2 highlights
the spiral arm in which the fragments form.}
\label{fig2c}
\end{figure*}

A close encounter of an intruder star with a massive disk may lead to
a direct ejection of fragments that have formed in the disk before
the encounter event. However, a fly-by of the intruder may also trigger ejection of 
dense elements of spiral arms essentially via the same multi-body gravitational interaction.
Figure~\ref{fig2c} illustrates this phenomenon and presents the surface density 
in a $4000 \times 4000$~AU$^{2}$ box 
centered around the target star in model~P4. The surface density is plotted at nine consecutive 
instances since
the launch of the intruder star. Arrow~1 points to the pre-existing massive fragment,
which is scattered and dispersed soon after the encounter with the intruder star,
likely because of strong gravitational perturbation and tidal torques during the close encounter.
At the same time, a dense spiral arm shown by Arrow~2 is ejected from the disk of the target star.
Two clumps form in the spiral arm close to each other and are ejected 
in the same direction. Due to the decreasing resolution
of the logarithmically spaced grid they merge at a distance of $> 3000$~AU from the central star.
However, a determination of the fragment masses, positions and velocities
at $\approx 3000$~AU shows that they are not bound to each other. 
A similar phenomenon was observed also in models~P4 and P6.

Table~\ref{table3} presents the properties of fragments formed in ejected spiral arms.
It is interesting to note that with masses ranging from 5 to 10~$M_{\rm Jup}$ 
and gas masses located inside the
Hill radius between 13 and 24~$M_{\rm Jup}$, the fragments formed in tidal arms 
are less massive than the fragments ejected from the disk. This is because they
form at a larger distance from the star, where the temperatures are lower,
corresponding to a lower Jeans mass. It is likely that the fragments will
accrete a fraction of the mass inside the Hill radius, though it is not clear
how much exactly. Therefore, the fragments will form either low-mass brown
dwarfs or free-floating planetary-mass objects.
The masses of ejected objects in our models
lie in a narrower mass range of 5--25~$M_{\rm Jup}$ (not taking the mass in the Hill radius into account)
and are skewed towards lower masses than what was found in
other numerical studies on the ejection phenomenon \citep[e.g.][]{Stamatellos2009,Thies2010,BV2012,Thies2015,Vor2016}. For instance, Thies et al. (2010,2015) reported the masses of ejected objects to lie in the 14--84~$M_{\rm Jup}$ range. \citet{Vor2016} found a similar
range of 21--145~$M_{\rm Jup}$, but shifted somewhat to higher masses. 
The reason for a narrower mass range in our study is that we considered encounters with only one disk
model and varied the initial impact parameters. We expect a better agreement once more disk models with
different preexisting characteristics of fragments are considered. Our models also seem to allow for
lower masses of ejected objects than normally found in other studies. This may be caused by either 
a limited range of parameter space in other works, as was already noted in Thies et al. (2010,2015), or by intrinsic properties of encounter-triggered ejections leading to tidal truncation of ejected 
fragments and allowing for ejection of entire spiral arms followed by fragmentation. We note 
that  our calculated masses can be brought to better agreement with other studies, if the mass in the Hill radius is added to the mass of ejected fragments. The fraction of the mass in the Hill radius
that will finally end up in the central object is however highly uncertain.

\begin{figure*}
\begin{centering}
\includegraphics[scale=0.9]{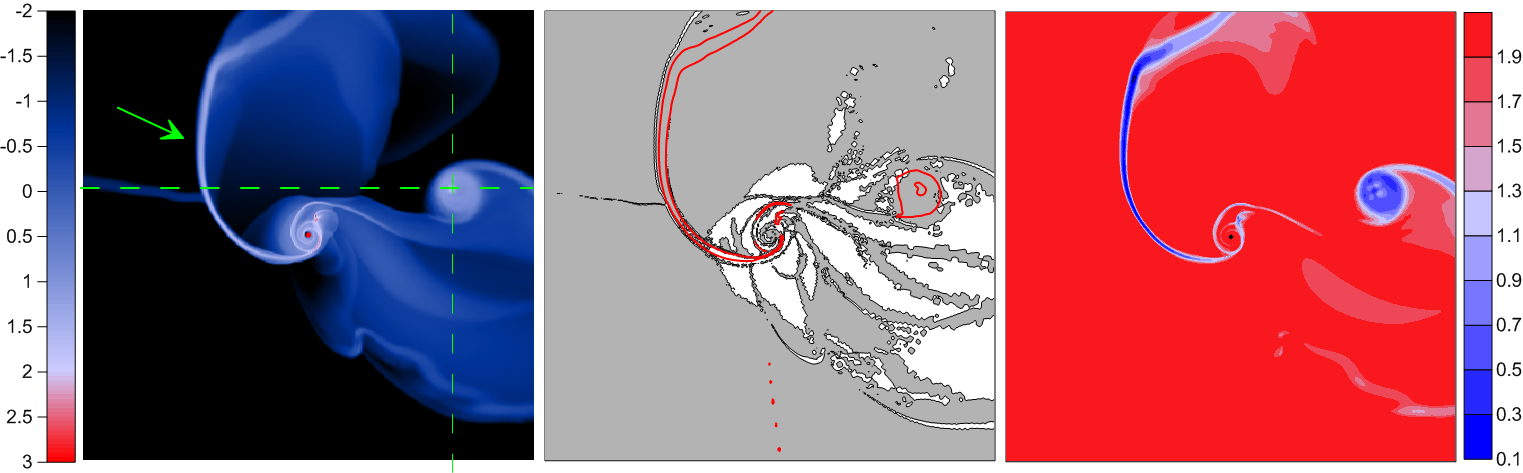} 
\par\end{centering}
\protect\protect\caption{{\bf Left}. Surface density of the accretion disk at $t=8000$~yr after the
launch of the intruder, just before the onset of fragmentation in the ejected spiral arm (shown by the
green arrow). Only the region inside 1000 AU from the central star is shown. The green dashed lines
mark the position of the intruder star. {\bf Middle}. Red
contour lines show the regions with the Toomre $Q$-parameter smaller than unity, while the grey-shaded
area indicated the disk regions where the Gammie ${\cal G}$-parameter is smaller than 3. {\bf Right}.
Ratio $N_{\rm Jeans}$ of the Jeans length to the minimum of the local grid cell size in 
both coordinate directions 
($r,\phi$), in log scale. The minimum value of $N_{\rm Jeans}$ in the ejected arm 
is about 2 and it increases to 3--5 in some parts of the arm,
meaning the the Truelove criterion may not be resolved in the entire spiral arm.}
\label{fig2d}
\end{figure*}

To check if the known fragmentation criteria are fulfilled in the ejected spiral arm, we
calculate the Toomre $Q$-parameter, the Gammie ${\cal G}$-parameter and the ratio $N_{\rm Jeans}$ 
of the Jeans length to the local grid spacing in a similar manner as was done in Section~\ref{sec:initialconditions}.
The results are shown in Figure~\ref{fig2d} for a time instance of $t=8000$~yr after the 
launch of the intruder, just before the spiral arm forms two
distinct fragments. Clearly, the ejected spiral arm is characterized by the Toomre parameter
$Q<1.0$ and by the Gammie parameter ${\cal G}<3.0$, fulfilling these fragmentation criteria. 
At the same time,  the minimum value of $N_{\rm Jeans}$ in the ejected arm 
is about 2 and it increases to 3--5 in some parts of the arm,
meaning the the Truelove criterion may not be resolved in the entire spiral arm.
The ejected spiral arms are colder than the disk and they have smaller Jeans lengths.
Therefore, numerical simulations with a higher numerical resolution are needed to 
fulfil the Truelove criterion in the entire ejected arm and
confirm that fragmentation in ejected spiral arms is not a numerical resolution effect. 

We note that all of the ejections of fragments formed in spiral arms happen in prograde
encounters. This is because prograde encounters form more extended
tidal structures, a behavior that has been observed in previous studies of
encounters as well (e.g. Shen et al., 2010; Forgan and Rice, 2009).

\section{Discussion and model caveats}
\label{discuss}

The formation mechanism of isolated brown dwarfs is traditionally attributed to 
the gravitational collapse of compact and dense pre-stellar cores, often 
complemented by some mechanisms that can 
stop accretion onto the central object before it can grow beyond the brown-dwarf mass regime
\citep{Whitworth2004,Padoan2004,Chabrier2011}. An alternative is the disk ejection mechanism when 
finished brown dwarf or their embryos 
are first formed in gravitationally unstable disks and then ejected into the intracluster medium 
\citep{Stamatellos2009, BV2012,Vor2016}. We note, however, that ejections due to internal multi-body gravitational interaction between the host star and fragments require at least two fragments 
(or finished brown dwarfs) to co-exist in the disk. These two fragments should also approach each
other quite closely, from several to tens of AU depending on their masses, 
to gain enough momentum and be ejected from the disk.

On the other hand, prograde encounters with intruder stars do not require a close approach 
for the fragment to be ejected. Our modeling shows that an approach of about 1000 AU
between the fragment and the intruder is sufficient for the fragment to be ejected (see model~P1).
We note that such encounters may be present
in the early disk evolution \citep[e.g.][]{Thies2010}, 
when disk gravitational fragmentation is also most likely \citep{VB2010}.
In addition, only one fragment is sufficient to pre-exist in the disk.
Both arguments increases the probability of fragment ejection via close encounters
with intruder stars. We note that those fragments that are not ejected directly as a result of
stellar encounter may end up on highly eccentric orbits with large semi-major axis of hundreds of AU.
These highly eccentric objects may be removed from the primary star later on through stellar-dynamical perturbations from passing stars within the birth cluster. Thus, even initially not ejected fragments or proto-brown dwarfs may end up being dissociated from their central star, increasing the net ejection
efficiency.


Finally, we want to mention several caveats to our modeling that need to be addressed
in the future studies. As was already mentioned in Section~\ref{spiral}, numerical 
simulations with higher resolution are needed to confirm that fragmentation in ejected spiral arms
is not a numerical resolution effect. In addition, the effect of accretion on the intruder star
and heating due to the intruder luminosity need to be addressed. In the current version, 
heating due to radiation of the host star is only included. 
Non-planar encounters need also to be studied. In this case, the gravitational force
acting from the intruder on the clump will torque the clump off the plane of the disk, potentially
leading to misaligned orbits of the clumps. Whether these clumps can still be ejected or rather 
form a population of bound but misaligned sub-stellar objects remains to be understood. 
Finally, a parameter study needs to be performed 
to assess the efficiency of fragment ejection depending on the mass of the intruder star
and  the efficiency of mass ejection by jets/outflows. As reported by \citet{Machida2012},
the fraction of mass ejected by jets and outflows may be higher (up to 50\%) than the
assumed 10\% in the present study. This may have a twofold effect. A higher ejection efficiency implies
a smaller stellar mass and a higher disk-to-star mass ratio, which may further increase the strength
of gravitational instability and stimulate more ejections. On the other hand, wide-angle outflows 
may reduce the available mass budget in the infalling envelope, causing lower mass
infall rates on the disk and weakening gravitational instability. Fully three-dimensional simulations
are needed to explore these complex feedback effects.

\section{Conclusions}
\label{conclude}
We simulated numerically a close encounter of a star plus disk system with a diskless intruder star.
Contrary to many previous studies of this phenomenon, we considered a massive disk
that has already undergone fragmentation due to gravitational instability in the embedded
phase of star formation. We followed
the evolution of the target system  for many orbital periods and considered both the
prograde and retorgrade co-planar encounters with a $1.2~M_\odot$ star. The initial
parameters of the target star plus disk system were set by a separate numerical hydrodynamics
simulation that was started from the gravitational collapse of a pre-stellar core with a mass
of $M_{\rm core}=1.08~M_\odot$ and was terminated at the end of the embedded phase when the mass 
of the host star was $M_\ast=0.63~M_\odot$ and the disk mass was $0.25~M_\odot$. 
Our findings can be summarized as follows.

\begin{itemize}

\item Close encounters of an intruder star with a target star hosting a massive, 
gravitationally unstable disk
can lead to the ejection of fragments that have formed in the disk of the target prior to collision.
In particular, prograde encounters are more efficient in ejecting the fragments than
the retrograde encounters, because the total angular momentum in the latter tends to cancel
out, making the ejection a less likely event. 

\item The masses of ejected fragments lie in the intermediate-mass brown-dwarf regime. The fragments
carry away a notable amount of gas in their radius of gravitational influence (the Hill sphere), 
implying that they can possess extended disks upon cooling and contraction to finished brown dwarfs.

\item Close prograde encounters can also lead to the ejection of entire spiral arms, followed by
fragmentation and formation of freely floating objects straddling the planetary mass limit. However,
numerical simulations with a higher resolution are needed to confirm this finding.


\end{itemize}

Stellar encounters can potentially present another mechanism for the formation of freely floating 
brown dwarfs and massive gaseous planets.
A crucial factor for the efficiency of the considered  mechanism is the frequency
of collisions with a massive and extended disk that has already undergone fragmentation.
These collisions must occur in the embedded phase of disk evolution or in the early T Tauri phase
when gravitational instability is strongest \citep[e.g.][]{VB2010,Tsukamoto2013}. This limits
the time period for collisions to the initial $\sim 0.5$~Myr of disk evolution. 
On the positive side, the encounter need not to be close. We found ejections with a periastron distance
of several hundreds of AU between the intruder and the target star. The intruder also need not to pass
close to the fragment -- ejections were found for the distance between the intruder star and the fragment of $\sim 1000$~AU.

Previous studies of the frequency of stellar encounters yielded results that 
depend on the cluster density \citep[e.g.][]{Olczak2010} and on the periastron distance.   
According to  \citet{Forgan2010}, close encounters in a cluster with number density 
100~pc$^{-3}$ and with a periastron
distance of a few tens of AU are rare and occur only for one out several thousand stars.
For more distant encounters of a few hundreds of AU, as in our studies, the probability of encounters
will increase by two orders of magnitude (the periastron distance enters as a power of 2), implying
one such encounter per several tens of stars.
\citet{Thies2010} estimated that an average 0.5 $M_\odot$ in an Orion-type cluster can experience 
an encounter with a periastron radius of
several hundred AU with a probability of $\sim10\%$, while being young and hosting a massive 
and extended disk.  Even with this probability of encounters, one ejection per 20 stars is inferred
(counting only prograde encounters), which is insufficient to explain the inferred ratio of brown 
dwarfs to stars -- one brown dwarf to 5--10 stars according to \citet{Luhman2007}. 
We note that these estimates does not take 
into account the internal ejection mechanisms due to close encounters between fragments 
in the disk \citep[e.g.][]
{Vor2016}. The total number of ejections per star may be higher if all ejection mechanisms are 
counted.  For instance,
\citet{Kroupa2003} based on the analysis of observational properties and distributions of brown
dwarfs reported the number of ejections to be about one per four stars (see also \citep{Thies2008}).

At the same time, this mechanism can produce freely floating sub-stellar objects with extended 
disks,  as also suggested by \citet{Thies2015}, because the ejected objects carry 
a notable amount of gas with high angular 
momentum around them. For instance, \citet[see also][]{Vor2016} reported that about half of
the ejected fragments are rotationally supported against gravity, implying the ratios of rotational
to gravitational energy on the order of unity.
Brown dwarfs formed via direct gravitational collapse can also have disks,
but their radius is expected to be smaller because pre-stellar cores have smaller angular 
momentum and are characterized by the ratio of rotational to gravitational energy 
not exceeding a few per cent \citep[e.g.][]{Caselli2002}. 
Therefore, brown dwarfs formed via ejection can be
observationally distinguished from the star-like mechanism of brown dwarf formation via 
direct gravitational collapse. 
The dust properties in these disks may also be different, featuring more grown dust
than their counterparts formed via direct collapse, because the fragments were formed
in the disk that may have already undergone a major episode of dust growth \citep{Vor2017}.

\section{Acknowledgments} 
We are thankful to the anonymous referee for useful suggestions that helped to improve the manuscript.
This work was supported by the Russian Science Foundation grant 17-12-01168.
The simulations were performed on the Vienna Scientific Cluster (VSC-2 and VSC-3)
and on the Shared Hierarchical Academic Research Computing Network (SHARCNET).
This publication (page charges) is supported by the Austrian Science Fund (FWF).

\end{document}